\documentclass[12pt,american,preprint, aps,nofootinbib]{revtex4-1}
\usepackage[T1]{fontenc}
\usepackage[latin9]{inputenc}
\usepackage{babel}
\usepackage{amsmath}
\usepackage{amssymb}
\usepackage{stmaryrd}
\usepackage{graphicx}
\usepackage[pdfusetitle,
 bookmarks=true,bookmarksnumbered=false,bookmarksopen=false,
 breaklinks=false,pdfborder={0 0 1},backref=false,colorlinks=false]
 {hyperref}

\makeatletter
\usepackage{etoolbox}
\usepackage{breqn}

\makeatletter
\let\cat@comma@active\@empty
\makeatother

\newcommand{\breqnoverloadothers}
{%
    \renewenvironment{equation}{\ignorespaces\begin{dmath}}{\end{dmath}\ignorespacesafterend}%
    \renewenvironment{equation*}{\ignorespaces\begin{dmath*}}{\end{dmath*}\ignorespacesafterend}%
    \renewenvironment{multline}{\ignorespaces\begin{dmath}}{\end{dmath}\ignorespacesafterend}%
    \renewenvironment{multline*}{\ignorespaces\begin{dmath*}}{\end{dmath*}\ignorespacesafterend}%

}

\newcommand\breqnundefineothers
{%
    \renewenvironment{equation}{}{}%
    \renewenvironment{equation*}{}{}%
    \renewenvironment{multline*}{}{}%

}

\AtBeginEnvironment{dmath}{\breqnundefineothers}
\AtBeginEnvironment{dmath*}{\breqnundefineothers}

\AtBeginEnvironment{dgroup}{\def\breqnundefineothers{}\breqnoverloadothers}
\AtBeginEnvironment{dgroup*}{\def\breqnundefineothers{}\breqnoverloadothers}

\newcommand\brwrap[3]{%
  \setbox0=\hbox{$#2$}
  \left#1\vbox to \the\ht0{\hbox to 0pt{}}\right.\kern-.2em
  \begingroup #2\endgroup\kern-.15em
  \left.\vbox to \the\ht0{\hbox to 0pt{}}\right#3
}

\makeatother

\begin{document}
\title{Holographic Reconstruction of Gravitational Perturbations in AdS/CFT
and Implications for Celestial Conformal Field Theory}
\author{David A. Lowe}
\author{Yiru Wang}
\author{Juanyi Yang}
\affiliation{Department of Physics, Brown University, Providence, RI 02912, USA}
\begin{abstract}
We begin by reexamining the holographic reconstruction of scalar fields
in four-dimensional anti-de Sitter spacetime, adopting a purely Lorentzian
signature derivation, reproducing earlier results of HKLL and generalizing
to arbitrary boundary metrics. The approach is extended to gravitational
perturbations, focussing on perturbations around $AdS_{4}$ and show
that the mapping can be formulated as a purely light-like integral
of the conformal field theory stress energy tensor. An example is
considered of relevance to the flat spacetime limit with nontrivial
BMS charges turned on, potentially providing a quantum field theory
definition of celestial CFT as a large central charge limit of a 3d
CFT.
\end{abstract}
\maketitle

\section{Introduction}

The conjectured mapping between conformal field theory and quantum
gravity in asymptotically anti-de Sitter spacetimes remains one of
the most promising avenues to provide a nonperturbative formulation
of string theory, and a definition of quantum gravity \citep{Maldacena:1999aa,Witten:1998qj}.
One of the goals of the present work is to provide a holographic reconstruction
of bulk gravitational perturbations around a four-dimensional anti-de
Sitter metric with a general conformal boundary metric. We formulate
the problem as a Lorentzian boundary value problem, avoiding boundary
constructions that require analytic continuation in the boundary coordinates.
To accomplish this, we revisit the scalar operator construction \citep{Hamilton:2005ju,Hamilton:2006fh,Hamilton_2006,Hamilton:2007wj,Kabat:2011rz},
and then generalize to spin 2 fields. In the spin 2 case, we show
the integral over boundary operators can be localized to the intersection
of the light-cone of the bulk point with the boundary at infinity.
This is to be contrasted with the spacelike region of integration
in the scalar case, and in the gravitational case studied in \citep{kabat}.

We then consider an example based on earlier work by Lowe and Ramirez
\citep{Lowe:2020qan}. There they studied collapsing/expanding spherical
gravitational waves in $AdS_{4}$. With Dirichlet boundary conditions
at infinity, such solutions represent gravitational waves that bounce
off the boundary at infinity. A boundary stress tensor source is present,
localized on the two-sphere of intersection. In the full nonlinear
solution studied by Lowe and Ramirez the solutions were characterized
by a single Virasoro algebra of charges, with non-vanishing central
charge. In these classical solutions, the boundary stress tensor does
not obey the dominant energy condition, having purely spacelike components.
However this violation of the dominant energy condition is instantaneous,
so does not appear to be ruled out. The discontinuity across the shocks
is characterized by a general holomorphic diffeomorphism of the two-sphere,
corresponding to a superrotation in the flat spacetime limit.

These shocks emerging from the boundary are closely connected with
corner conditions. In the formulation of the initial value problem
in AdS, imposing Dirichlet boundary conditions on the metric at spatial
infinity, leads to similar shocks emerging from the intersection of
the initial value surface with infinity. The matching conditions have
been studied in \citep{Friedrich:1995vb,Horowitz:2020aa}.

The present setup offers a framework where the fully quantum holographic
map may be defined, as at least perturbatively, one may represent
any quantum superposition of gravitons as a state in the conformal
field theory, and use the CFT to generate a unitary time evolution.
It is natural to conjecture this provides a non-perturbative definition
of quantum gravity, both with AdS asymptotic conditions, and in the
asymptotically flat limit \citep{Pasterski:2021aa}. Thus the holographic
dual of gravity under these conditions would be a 3d CFT. The present
paper shows how to reconstruct general bulk perturbations from this
CFT data.

\section{Holographic Reconstruction of Scalar Fields}

We begin by revisiting the holographic reconstruction of massive scalar
fields in anti-de Sitter spacetime, generalizing previous results
to general boundary metrics, and emphasizing a purely Lorentzian signature
approach. The results will then be much more easily applied to realistic
boundary value problems. We work with an AdS metric in Fefferman-Graham
form \citep{fefferman}
\begin{equation}
ds^{2}=\frac{dz^{2}}{z^{2}}+\frac{g_{ij}(x,z)}{z^{2}}dx^{i}dx^{j}\label{eq:fefferman}
\end{equation}
where $i,j=0,\cdots,2$ run over the transverse directions of $AdS_{4}$.
We will adopt the notation $X^{\mu}$ to denote the coordinates $(x^{i},z)$
with $\mu=0,\cdots,3$ and $g_{\mu\nu}^{(4)}$ to denote the 4-dimensional
metric.

Our goal is to express the a general solution to the bulk wave equation
\[
\left(\oblong-m^{2}\right)\phi(X)=0
\]
 in terms of boundary data, using Green's theorem. To accomplish this
we need the Lorentzian Green function

\begin{equation}
(\Box-m^{2})G(X,X')=\frac{1}{\sqrt{-g^{(4)}}}\delta^{(4)}(X|X').\label{eq:massivescalar}
\end{equation}
We seek a solution to this equation which vanishes at timelike separations,
but will be non-vanishing for null and spacelike separations. This
condition will force the boundary behavior of the Green function to
be non-normalizable, however this will be precisely what we need to
then apply Green's theorem to obtain the holographic construction.
The Green function can be expressed in terms of the invariant distance
between the points $x$ and $x'$ . We parameterize this by $\sigma(x,x')$,
which is most simply defined in embedding space (i.e. signature $(2,3)$
Minkowski spacetime) coordinates $W$ and $W'$ as

\begin{equation}
\sigma=\frac{1}{2}(W-W')^{2}\label{eq:invdist}
\end{equation}
with $W^{2}=-W_{0}^{2}-W_{1}^{2}+W_{2}^{2}+W_{3}^{2}+W_{4}^{2}$,
and work in units where the AdS radius of curvature is $1$, so that
for a point $X$ on AdS, $W^{2}=-1$. The Green function then takes
the form

\[
G(\sigma)=-\frac{1}{8\pi}\left(\left(\frac{2+m^{2}}{2}\right)\,_{2}F_{1}\left(3-\Delta,\Delta,2,-\frac{\sigma}{2}\right)\Theta(\sigma)+\delta(\sigma)\right)
\]
where $\Delta=\frac{3}{2}+\sqrt{\frac{9}{4}+m^{2}}$ which ends up
being the conformal weight of the primary operator of the boundary
CFT. The coefficient of the $\delta(\sigma)$ term is fixed by the
coefficient of the $\delta^{(4)}(X,X')$ in \eqref{eq:massivescalar}.
This is turn sets the coefficient of the $\theta(\sigma)$ term in
the limit $\sigma\to0$. The Green function is then uniquely determined
by the condition that it vanish at spacelike separations. The distributional
nature of the Lorentzian Green function is often sidestepped in the
literature by beginning in Euclidean signature, and then Wick rotating.
This approach is advantageous for working in general dimensions, but
for the purposes of the present paper we desire to avoid that so we
can formulate the problem as a genuine boundary value problem.

The massless scalar Green function is particularly simple, with $m^{2}=0$
the above reduces to
\begin{equation}
G(\sigma)=-\frac{1}{8\pi}\left(\theta(\sigma)+\delta(\sigma)\right)\label{eq:scalargreen}
\end{equation}
The $\delta(\sigma)$ term is familiar from the electromagnetic potential
calculations of standard texts such as \citep{jackson}. Coupling
to the curvature of AdS induces the subleading distributional term
$\theta(\sigma)$ with a coefficient fixed by that of the $\delta(\sigma)$
term. 

\subsection{Smearing function}

We now apply Green's second identity
\begin{align}
\int_{M}d^{3}X'\sqrt{-g^{(4)}}\,\left(\phi(X')\,(\Box-m^{2})\,G(X;X')-G(X,X')(\Box-m^{2})\phi(X')\right)\nonumber \\
=\int_{\partial M}d^{3}S'{}^{\mu}\left(\phi(X')\frac{\partial G(X,X')}{\partial n'{}^{\mu}}-G(X,X')\frac{\partial\phi(X')}{\partial n'{}^{\mu}}\right)\label{eq:greentwo}
\end{align}
where the region $M$contains the point $X$ and extends outward to
$z=\epsilon$. We insist that $\phi$ be a normalizable perturbation
which implies a falloff of
\[
\phi(x,z)\sim z^{\Delta}\phi_{0}(x)\,.
\]
One may then evaluate the right hand side of \eqref{eq:greentwo}
and rearrange to obtain
\[
\phi(X)=\int_{\partial M}d^{3}x'\sqrt{-g^{(0)}\,}K(X,x')\phi_{0}(x')
\]
where the boundary metric is $g_{ij}^{(0)}(x)=\lim_{z\to0}g_{ij}(x,z)$
and the smearing function $K$ is vanishing at timelike separations
\[
K(X,x')=-\frac{2^{\Delta-4}\Gamma(\Delta-\frac{1}{2})}{\pi^{3/2}\Gamma(\Delta-2)}\tilde{\sigma}(X,x')^{\Delta-3}\theta(\tilde{\sigma}(X,x'))
\]
where we have defined $\tilde{\sigma}(X,x')=\lim_{z'\to0}z'\sigma(X,x')$. 

In the massless case, where $\Delta=3$ we have
\[
K(X,x')=-\frac{3}{8\pi}\theta(\tilde{\sigma}(X,x'))
\]
and the smearing function is simply a constant, within the spacelike
wedge.

\section{Holographic Reconstruction of Gravitational Perturbations}

In this section our aim is to generalize the above results to spin
2 perturbations around $AdS_{4}$, namely extending previous results
to a general boundary metric and avoid the use of analytic continuation
in the boundary fields to build the holographic map \citep{kabat}.
Many of the relevant formulas have been collected in \citep{DHoker:1999bve},
so we begin by briefly summarizing those results. We define the perturbation
of the metric as $h_{\mu\nu}$ with 
\[
ds^{2}=g_{\mu\nu}^{(4)}dX^{\mu}dX^{\nu}=\frac{dz^{2}}{z^{2}}+\frac{g_{ij}^{(0)}(x,z)}{z^{2}}dx^{i}dx^{j}+h_{\mu\nu}(x,z)dX^{\mu}dX^{\nu}
\]
in Fefferman-Graham gauge \ref{eq:fefferman}. Here $g_{ij}^{(0)}(x,z)$
is part of the $AdS$ metric, while $g_{ij}^{(0)}(x)=\lim_{z\to0}g_{ij}^{(0)}(x,z)$
is the boundary metric. We work in transverse-traceless gauge to isolate
the physical gravitational wave perturbations, $h_{\mu}^{\mu}=0$
and $\nabla^{\mu}h_{\mu\nu}=0$. Later we will need to further impose
Fefferman-Graham gauge, which requires $h_{zi}=h_{zz}=0$, which we
consider in the subsequent section. The linearized Einstein equation
becomes
\[
\left(\nabla^{\sigma}\nabla_{\sigma}+2\right)h_{\mu\nu}=-2T_{\mu\nu}
\]
where we impose the condition $T_{\mu}^{\mu}=0$, since we will ultimately
be interested in a conformally invariant boundary source. To solve
this equation around for a general choice of AdS metric and general
boundary conditions (compatible with the above conditions), it is
convenient to use the maximally symmetric bitensor formalism of \citep{Allen:1985wd}
which was adopted in \citep{DHoker:1999bve}. In the case at hand,
we use the invariant distance $\sigma$ of \eqref{eq:invdist}. As
pointed out in \citep{DHoker:1999bve}, there is one universal piece
of the graviton propagator which propagates the physical degrees of
freedom, and the remainder correspond to different gauge fixing conditions.
In general there is also a trace component, but we will not need that
if we impose the condition $T_{\mu}^{\mu}=0$. Therefore we will only
need one of the five tensor structures possible, and proceed to check
this satisfies the necessary conditions. The bulk graviton Green function
is then
\begin{equation}
G_{\mu\nu;\mu'\nu'}(X,X')=\left(\partial_{\mu}\partial_{\mu'}\sigma\partial_{\nu}\partial_{\nu'}\sigma+\partial_{\mu}\partial_{\nu'}\sigma\partial_{\nu}\partial_{\mu'}\sigma\right)G(\sigma)\label{eq:ansatz}
\end{equation}
which is to satisfy the equation
\begin{equation}
\left(\nabla^{\lambda}\nabla_{\lambda}+2\right)G_{\mu\nu;\mu'\nu'}=-\left(g_{\mu\mu'}g_{\nu\nu'}+g_{\mu\nu'}g_{\nu\mu'}-g_{\mu\nu}g_{\mu'\nu'}\right)\delta^{(4)}(X-X')/\sqrt{-g^{(4)}}+\nabla_{\mu'}\Lambda_{\mu\nu;\nu'}+\nabla_{\nu'}\Lambda_{\mu\nu;\mu'}\label{eq:greeneqn}
\end{equation}
where $\Lambda_{\mu\nu;\mu'}$ represents a diffeomorphism in the
$X'$ coordinates that should vanish when integrated against a conserved
stress tensor. The parallel propagator is 
\[
g_{\mu\nu'}=-\partial_{\mu}\partial_{\nu'}\sigma+\frac{\partial_{\mu}\sigma\partial_{\nu'}\sigma}{\sigma+2}
\]
(correcting a typo in \citep{DHoker:1999bve}). It is shown in \citep{DHoker:1999bve}
that the scalar $G(\sigma)$ satisfies the massless scalar wave equation
of the previous section. We will follow the strategy there of choosing
boundary conditions that give the solution \eqref{eq:scalargreen}.

We then build a bulk perturbation using
\[
h_{\mu\nu}(X)=\int d^{4}X'\sqrt{-g^{(4)}}G_{\mu\nu;\mu'\nu'}(X,X')T^{\mu'\nu'}(X')
\]
but will localize the source on the boundary (taking a limit $\epsilon\to0)$
using
\begin{equation}
T^{\mu'\nu'}(X')=3\delta(z'-\epsilon)z'{}^{6}T_{(b)}^{\mu'\nu'}(x')\label{eq:bstress}
\end{equation}
and identify $T_{(b)}^{\mu'\nu'}(x')$ with the boundary stress tensor,
which we take to satisfy the conditions $T_{(b)}^{z\mu}=0$, $T_{(b)\mu}^{\mu}=0$
and $\nabla_{\mu}T_{(b)}^{\mu\nu}=0$ which in Fefferman-Graham coordinates
also implies $\nabla_{i}T_{(b)}^{ij}=0$ and $T_{(b)i}^{i}=0$ with
respect to the boundary metric. The factor of $3$ will be discussed
later in the section. Plugging in the solution \eqref{eq:ansatz}
gives
\begin{align}
h_{\mu\nu}(X) & =\lim_{\epsilon\to0}3\int d^{3}x'\sqrt{-g^{(0)}}\left(\partial_{\mu}\partial_{\mu'}\bar{\sigma}\partial_{\nu}\partial_{\nu'}\bar{\sigma}+\partial_{\mu}\partial_{\nu'}\bar{\sigma}\partial_{\nu}\partial_{\mu'}\bar{\sigma}\right)G(\sigma)T_{(b)}^{\mu'\nu'}(x')\nonumber \\
 & =-\frac{3}{4\pi}\int_{\Sigma(X)}d^{3}x'\sqrt{-g^{(0)}}\partial_{\mu}\partial_{\mu'}\bar{\sigma}\partial_{\nu}\partial_{\nu'}\bar{\sigma}T_{(b)}^{\mu'\nu'}(x')\label{eq:3dsmear}
\end{align}
where $\bar{\sigma}=\lim_{\epsilon\to0}\epsilon\sigma|_{z'=\epsilon}$
is finite in this limit and $\Sigma(X)$ is the region on the boundary
spacelike separated from point $X$. In the $\epsilon\to0$ limit
the $\delta(\sigma)$ term in \eqref{eq:scalargreen} does not contribute
to this integral.

To proceed, we note that $\nabla_{\mu'}\partial_{\nu'}\partial_{\mu}\sigma=g_{\mu'\nu'}\partial_{\mu}\sigma$
which implies that the vector $\xi_{\mu'}=\partial_{\mu}\partial_{\mu'}\bar{\sigma}$
is a conformal Killing vector. We may therefore use conservation of
$T_{(b)}^{\mu'\nu'}(x')$ and vanishing trace to integrate by parts
with respect to $\partial/\partial x^{\mu'}$. The boundary of $\Sigma(X)$
is the intersection of the boundary with the light-cone of point $X$,
which we denote by $\Omega(X)$. The integral then reduces to
\[
h_{\mu\nu}(X)=-\frac{3}{4\pi}\int_{\Omega(X)}d^{2}x'\sqrt{-g^{(2)}}\partial_{\mu}\bar{\sigma}\partial_{\nu}\partial_{\nu'}\bar{\sigma}\eta_{\mu'}T_{(b)}^{\mu'\nu'}(x')
\]
where $\eta_{\mu'}$ is a unit timelike vector normal to the spacelike
surface $\Omega(X)$ and $g_{\mu\nu}^{(2)}$ is the induced metric
on the surface.

Next we examine the components $h_{z\mu}(X)$. We can write the result
as
\begin{align}
h_{\mu z}(X) & =-\frac{3}{4\pi}\int_{\Omega(X)}d^{2}x'\sqrt{-g^{(2)}}\left(\partial_{\mu}\partial_{\nu'}\bar{\sigma}\partial_{z}\bar{\sigma}\right)\eta_{\mu'}T_{(b)}^{\mu'\nu'}(x')\nonumber \\
 & =-\frac{3}{4\pi}\int_{\Omega(X)}d^{2}x'\sqrt{-g^{(2)}}\left(\partial_{\mu}\partial_{\nu'}\bar{\sigma}\right)f(z)\eta_{\mu'}T_{(b)}^{\mu'\nu'}(x')\label{eq:hzmu}
\end{align}
where in the second line we have evaluated $\partial_{z}\bar{\sigma}=f(z)$
on $\Omega(X)$ and note it is independent of position in the transverse
space. Again we note that $\left(\partial_{\mu}\partial_{\nu'}\bar{\sigma}\right)$
is a conformal Killing vector so the second line is a conserved quantity
when integrated against a conserved and traceless stress tensor. Note
that $\Omega(X)$ is composed of a past and future branch of the light-cone,
so provided the charge associated with this conformal Killing vector
is conserved, each contribution will cancel in the integral, and $h_{\mu z}(X)=0$.
Thus choosing a traceless conserved boundary stress tensor, with conservation
of the charge
\begin{equation}
Q_{\mu}=\int d^{2}x'\sqrt{-g^{(2)}}\left(\partial_{\mu}\partial_{\nu'}\bar{\sigma}\right)\eta_{\mu'}T_{(b)}^{\mu'\nu'}(x')\label{eq:physcond}
\end{equation}
when integrated across a null (or spacelike) boundary hypersurface,
implies the perturbation remains in Fefferman-Graham gauge. 

Now, let us prove $Q_{\mu}$ is conserved by beginning with the equation
\[
\int_{V}d^{3}x'\sqrt{-g^{(0)}}\nabla_{j'}\left(\xi_{i'}T_{(b)}^{i'j'}\right)=0
\]
which is true if $\xi_{i'}$ is a conformal Killing vector and $T_{(b)}$
is conserved and traceless on the boundary. Here $\nabla_{j'}$ is
defined with respect to the boundary metric and $V$ is a 3d region
on the boundary extended in the time direction. We will get a different
$\xi_{i'}$ for each choice of $\mu$, but it is convenient to drop
this index for the moment. Now integrate by parts with respect to
$x^{j'}$ and we find
\[
\int_{\partial V}d^{2}x'\sqrt{-g^{(2)}}\xi_{i'}\eta_{j'}T_{(b)}^{i'j'}=0
\]
so conservation of $T_{(b)}$ and vanishing trace implies $Q_{\mu}$
is conserved (taking the boundary of $V$ to be boundary hypersurfaces).

For the transverse components we then have
\begin{equation}
h_{ij}(X)=-\frac{3}{4\pi}\int_{\Omega(X)}d^{2}x'\sqrt{-g^{(2)}}\partial_{i}\bar{\sigma}\partial_{j}\partial_{j'}\bar{\sigma}\eta_{i'}T_{(b)}^{i'j'}(x')\label{eq:fans}
\end{equation}
which is the final form for the bulk perturbation.

\subsection{Comparison to previous results}

Previous results were written as a 3d integral over the boundary for
the case of conformally flat coordinates \citep{kabat}. In that case
we find we can evaluate \eqref{eq:3dsmear} and find
\begin{align*}
h_{ij}(X) & =-\frac{3}{8\pi}\int_{\Sigma(X)}d^{3}x'\sqrt{-g^{(0)}}\left(\partial_{i}\partial_{i'}\bar{\sigma}\partial_{j}\partial_{j'}\bar{\sigma}+\partial_{i}\partial_{j'}\bar{\sigma}\partial_{j}\partial_{i'}\bar{\sigma}\right)T_{(b)}^{i'j'}(x')\\
 & =-\frac{3}{4\pi z^{2}}\int_{\Sigma(X)}d^{3}x'\,T_{(b)ij}(x')
\end{align*}
If we proceed to analytically continue the boundary spatial coordinates
to imaginary values (as well as in the boundary metric $g_{ii'}^{(0)}$),
this reproduces formula (37) in \citep{kabat}. The factor of 3 in
our definition of the boundary stress-energy tensor \eqref{eq:bstress}
ensures the matching of the coefficients.

\section{BMS Style Charges}

\begin{figure}
\includegraphics[scale=0.35]{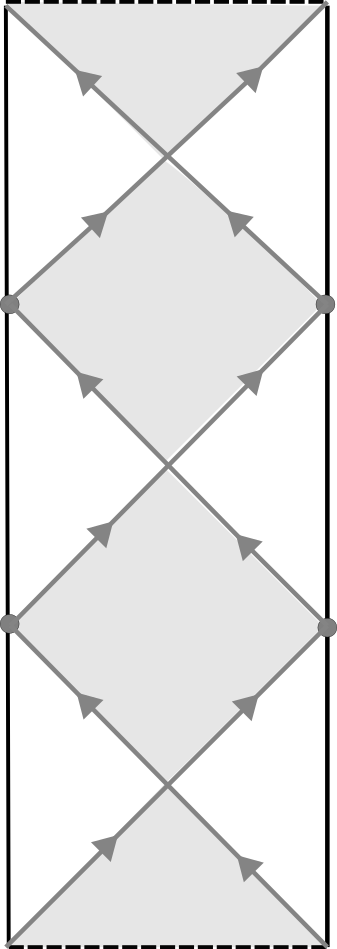}\caption{Bouncing gravitational shocks in asymptotically AdS spacetime. The
shaded regions show causal diamonds that become asymptotically flat
spacetimes in limit of vanishing cosmological constant. The shaded
dots represent boundary stress energy that generates BMS superrotations
in this limit. The general gravitational perturbation is holographically
reconstructed by a traceless, conserved stress energy tensor on the
full 3d boundary.\protect\label{fig:Bouncing-gravitational-shocks}}

\end{figure}

The present construction allows one to build a general gravitational
perturbation around the background studied in \citep{Lowe:2020qan}.
In this solution the boundary metric is built out of a sequence of
$dS_{3}$ patches and is illustrated in figure \ref{fig:Bouncing-gravitational-shocks}.
Within white regions, the metric takes the form
\begin{equation}
ds^{2}=\frac{dz^{2}}{z^{2}}+\frac{1}{16z^{2}}(1-z^{2})^{2}\left(\left(e^{t/2}+e^{-t/2}\right)^{2}d\Omega^{2}-dt^{2}\right)\,.\label{eq:ds3slice}
\end{equation}
This corresponds to a $dS_{3}$ slicing of $AdS_{4}$ where the boundary
now has topology $S_{2}\times R$ rather than the $R^{3}$ of the
conformally flat slicing. We note the background metric is in Fefferman-Graham
gauge, with
\begin{align*}
g_{ij} & =g_{ij}^{(0)}+z^{2}g_{ij}^{(2)}+\cdots\\
g_{ij}^{(0)} & =\frac{1}{16}\left(\left(e^{t/2}+e^{-t/2}\right)^{2}d\Omega^{2}-dt^{2}\right)
\end{align*}
and the invariant distance is
\begin{align*}
\sigma & =\frac{1}{2}(W-W')^{2}\\
 & =\frac{2(z-z')^{2}+(z^{2}-1)(z'{}^{2}-1)\left(1+\sinh\frac{t}{2}\sinh\frac{t'}{2}-\cosh\frac{t}{2}\cosh\frac{t'}{2}\left(\cos\left(\phi-\phi'\right)\sin\theta\sin\theta'+\cos\theta\cos\theta'\right)\right)}{4zz'}\,.
\end{align*}
One then has all the ingredients needed to evaluate the general gravitational
perturbation using \eqref{eq:fans}. 

The final form for our generic gravitational perturbation \eqref{eq:fans}
is only sensitive to the boundary stress tensor on the intersection
of the null cone with the boundary. However in the classical solutions
constructed in Lowe-Ramirez, the boundary stress tensor was only non-vanishing
at particular times, where a nontrivial 2d spatial boundary stress
tensor could appear, as shown by the gray dots in figure \eqref{fig:Bouncing-gravitational-shocks}.
While these solutions are valid classically, it is somewhat unclear
as to whether they arise from well-behaved quantum states in the CFT,
since the dominant energy condition is violated. In particular, the
boundary stress energy in the solution of Lowe-Ramirez is non-vanishing
only at $t=0$ in the patch \eqref{eq:ds3slice}, but has purely spacelike
components, so still satisfies the local conservation and traceless
conditions. 

The coordinate patch \eqref{eq:ds3slice} only covers half of global
AdS, so if one wishes to do describe the universal cover of AdS (unwrapping
the time direction) one needs a periodic array of such patches, together
with a corresponding periodic array of boundary sources as shown in
figure \ref{fig:Bouncing-gravitational-shocks}. The upshot is that
points spacelike separated from the boundary stress tensor localized
at $t=0$, as shown in the gray causal diamonds in figure \ref{fig:Bouncing-gravitational-shocks},
are lightlike separated from the neighboring image sources. In applying
\eqref{eq:fans} one must include all the relevant boundary sources
lightlike separated from the bulk point in question. Taken together,
one then has a formula for the bulk reconstruction of a general metric
perturbation for a general stress energy tensor. With a coordinate
transformation to complex coordinates on the two-sphere, as explained
in \citep{Lowe:2020qan}, the 2d stress tensor is a sum of holomorphic
and antiholomorphic components, as is typical in 2d CFT. One may then
take a flat-spacetime limit where the metric perturbation remains
finite, to obtain a holographic reconstruction of the bulk metric. 
\begin{acknowledgments}
The research of D.L. is supported in part by DOE grant de-sc0010010.
\end{acknowledgments}

\bibliographystyle{utcaps}
\bibliography{holograv}

\end{document}